\documentclass{elsart}
\usepackage{epsfig}

\begin{document}

\begin{frontmatter}
\title{A Simple Parameterization of the Cosmic-Ray Muon Momentum Spectra at the Surface as a Function of Zenith Angle}

\author{D. Reyna}
\address{HEP Division, Argonne National Laboratory, 9700 S. Cass Ave.,
Argonne, IL 60439, USA}

\begin{abstract}
The designs of many neutrino experiments rely on calculations of the background
rates arising from cosmic-ray muons at shallow depths.  Understanding the 
angular dependence of low momentum cosmic-ray muons at the surface is
necessary for these calculations.  Heuristically, from examination of the data,
a simple parameterization is proposed, based on a straighforward scaling 
variable.  This in turn, allows a universal calculation of the differential 
muon intensity at the surface for {\em all} zenith angles and essentially all 
momenta.
\end{abstract}

\end{frontmatter}

\section{Introduction}
There is a need among the experimental design community for the ability to 
accurately predict backgrounds from cosmic ray muons at relatively shallow 
depths -- less than a few hundreds of meters of water equivalent (m.w.e.).  
In the past, most studies have focused on muon fluxes and intensities at very 
deep sites (greater than 1 km.w.e.) which do not accurately predict the
behavior of muons below a few hundred GeV.  Since shallow sites are dominated
by muons between 3--20 GeV, the characteristic distributions of these muons
can have a great impact on the designs of experimental laboratories and 
shielding geometries.  

The momentum distribution of the vertical muon intensity 
($I_{\textrm v}(p_{\mu})$ [cm$^{-2} \cdot$ sr$^{-1} \cdot$ s$^{-1} \cdot$ GeV$^{-1}$]) 
at the surface is fairly well known, as many experiments have provided 
measurements.  
However, when calculating the total rate of muons for a given shielding 
configuration, it is the interplay between the angular distribution of muons
as a function of momentum and the angular distribution of shielding and
overburden which
is important.  While it is generally accepted\cite{PDG} that the muon 
angular distribution is $\propto \cos^2\theta$ for low muon momenta of 
around 3 GeV, and that at higher momenta ($p_\mu >$ 100--200 GeV) it 
approaches 
a $\sec\theta$ distribution (for $\theta < 70^\circ$), there is currently 
no simple way to accurately estimate the muon momentum spectra over 
{\em all} angles.

In this work, experimental data in which muon momentum distributions have 
been recorded at various zenith angles have been compared.  A simple 
relationship has been found, 
$I(p_\mu,\theta) = \cos^3(\theta)I_{\textrm v}(p_\mu \cos\theta)$ 
(cf. Fig.~\ref{F-bestFit}).  It relates all angles to the differential 
vertical muon intensity distribution $I_{\textrm v}$ by simply scaling 
the independant variable (momentum) by $\cos\theta$ and the dependant
variable ($I$) by $\cos^3\theta$.  The theoretical
interpretation of this intriguing universality remains unclear.  
An improved fit is provided which should allow surface muon intensity 
predictions over all zenith angles for the momentum range of
most relevance to shallow sites: $p_\mu > 1$GeV and 
$p_\mu < 2000$GeV/$\cos\theta$.

\section{Data Selection}

Before analysing the data, some brief comments on the selection criteria
are required.  An attempt was made to include as much surface muon data
at various zenith angles as possible.  A majority of the available cosmic-ray
muon data was recorded at underground locations.  Since these data are usually
reported after slant-depth corrections which inherently assume certain
zenith angle dependancies, it was decided to exclude all data recorded at
depth.  Additionally, surface experiments for which the angular acceptance 
was not clearly specified or the systematic errors were not discussed were 
not included in this study.

The selected data, shown in Table~\ref{T-exp}, come from six surface muon 
experiments which measured intensity as a function of muon momentum and 
zenith angle.  They span zenith angles from the vertical to the horizontal
and cover muon momenta up to ~2000 GeV/$\cos\theta$.  

\begin{table}[htb]
\caption{The six experimental data sets that were included in this study.  For
each experiment, the ranges in energy and angular acceptance are shown
for the individual data sets.  When comparing data sets, the average value of 
$\cos(\theta)$ was used.} \label{T-exp}
\begin{center}
\begin{tabular}{|l c|c|c|}
\hline
Experiment & 
  & Zenith Angle Range ($^\circ$)& $p_\mu$ (GeV) \\
\hline \hline
Nandi and Sinha\cite{Nandi}  
  & 0$^\circ$ & 0 -- 0.3     & 5 -- 1200 \\ \hline
MARS\cite{Ayre}              
  & 0$^\circ$ & 0 -- 0.08    & 20 -- 500 \\ \hline
Kellogg et al.\cite{Kellogg} 
  & 30$^\circ$ & 25.9 -- 34.1 & 50 -- 1700 \\ 
  & 75$^\circ$ & 70.9 -- 79.1 & 50 -- 1700 \\ \hline
OKAYAMA\cite{Tsuji}  
  & 0$^\circ$ & 0 -- 1       & 1.5 -- 250 \\
  & 30$^\circ$ & 26 -- 34     & 2 -- 250 \\
  & 60$^\circ$ & 59 -- 61     & 3 -- 250 \\
  & 75$^\circ$ & 69 -- 81     & 3 -- 250 \\
  & 80$^\circ$ & 79 -- 81     & 3 -- 150 \\ \hline
Kiel-Desy\cite{Jokisch}
  & 75$^\circ$ & 68 -- 82     & 1 -- 1000 \\ \hline
MUTRON\cite{Matsuno}   
  & 89$^\circ$ & 86 -- 90     & 100 -- 20,000 \\
\hline
\end{tabular}
\end{center}
\end{table}

A few comments should be made about how the data are
included here.  Whenever a systematic error was listed, but not
already included in the tabulated data, it was combined in quadrature with
the statistical errors.  The Kiel-Desy experiment reported differences 
between their measured differential spectra and the phenomenological fit
used for correcting the data, primarily
at low momenta (1--20 GeV).  These differences were included as an additional 
systematic error. 

For the Kellogg data, the angular acceptance of the magnetic spectrometer
is listed as $\pm 11.4^\circ$.  However, the efficiencies fall off rapidly
and it is stated that the data are heavily peaked in the region of 
$\pm 4.1^\circ$ from the central zenith angle.  For that reason, the 
lower angular range was used in this comparison.  

There is some duplication in the OKAYAMA data set which should also be 
mentioned.  The angular acceptance of the OKAYAMA telescope was $\pm 1^\circ$.
The data at 30$^\circ$ and 75$^\circ$ were combined from smaller sets
of zenith angle data to be more comparable to other data sets ({\em e.g.}
those from Kellogg et al. or Kiel-Desy).  As a result, the data at 80$^\circ$
are a subset of the 75$^\circ$ data set.  It was felt, however, that
including that data was still relevant to understanding the zenith angle
dependence.

\section{Data Comparison}
\begin{figure}[htb]
\setlength{\unitlength}{1cm}
\begin{center}
\epsfig{file=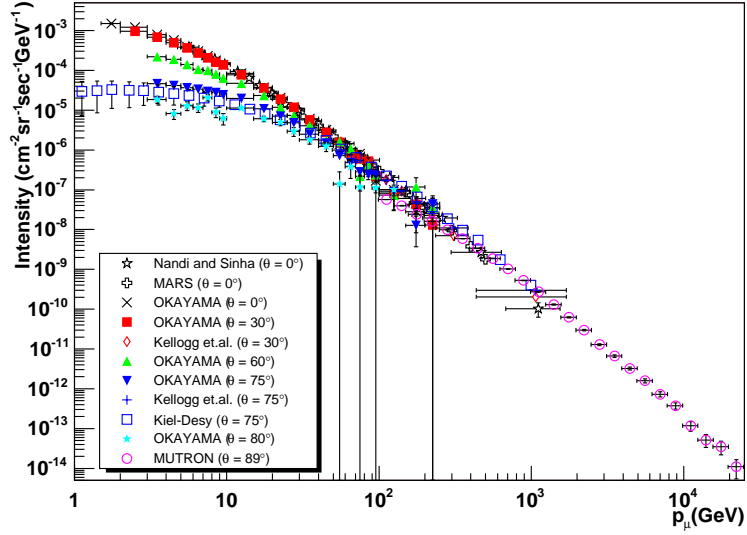, width=0.8\textwidth}
\end{center}
\caption{The differential surface muon intensity is plotted as a function 
of muon momentum for all the data sets listed in Table~\ref{T-exp}.}
\label{F-allData}
\end{figure}

In Fig.~\ref{F-allData}, the differential muon intensity data from all of 
the data sets listed in Table~\ref{T-exp} are plotted together as a 
function of the muon momentum.  One notes the power law dependence of
the higher momentum data that can be approximated by $p_\mu^{-3.7}$.  Attempts
to describe the low momentum fall off and angular dependence as simple 
corrections to the primary power law have had minimal success.  

\begin{figure} [htb]
\setlength{\unitlength}{1cm}
\begin{center}
\epsfig{file=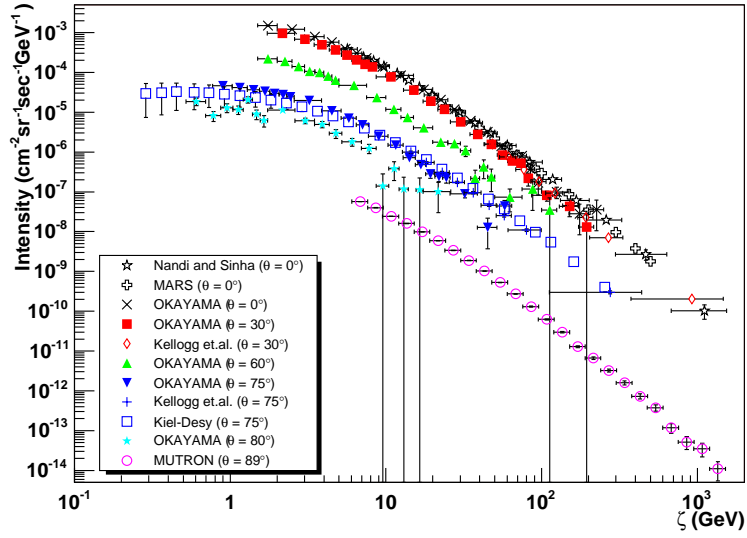, width=0.8\textwidth}
\end{center}
\caption{The differential surface muon intensity is plotted as a 
function of $\zeta$ as defined in Eq.~\ref{E-zeta}.  One clearly notices 
the similarity of spectral shape and the grouping of the different 
data by zenith angle.}
\label{F-zetaData}
\end{figure}

However, a similarity in spectral shape can be seen between all the data sets 
(Fig.~\ref{F-zetaData}) if a simple scaling variable is introduced
\begin{equation}
\zeta = p_\mu \cos(\theta). 
\label{E-zeta}
\end{equation}

\begin{figure}[htb]
\setlength{\unitlength}{1cm}
\begin{center}
\epsfig{file=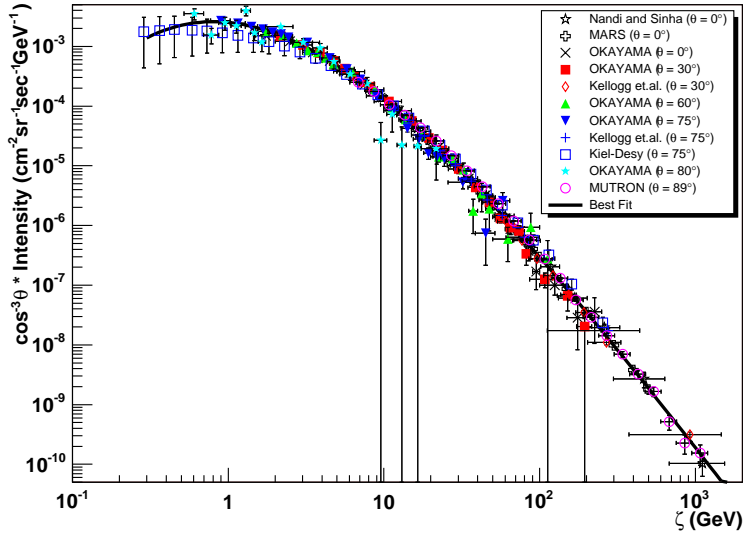, width=0.8\textwidth}
\end{center}
\caption{Surface muon intensity is plotted as a function of 
$\zeta = p_\mu \cos(\theta)$.  Each data set is scaled by a factor
$1/\cos^3(\theta)$. Also shown is the best fit, derived from combining 
Eqs.~\ref{E-intensity} and \ref{E-bugaev} as described in 
Section~\ref{S-models}.} \label{F-bestFit}
\end{figure}

Based on this observation, a succesful attempt was made to find a scale 
factor for all momenta $\propto1/\cos^n(\theta)$. Optimal
agreement was found at a value of $n = 3$ (shown in Fig.~\ref{F-bestFit}).  

This implies that a simple relationship exists in the data that relates
the muon intensity at any momentum and angle to the 
vertical intensity ($I_{\textrm v}$) by 
\begin{equation}
I(p_\mu,\theta) = \cos^3(\theta)I_{\textrm v}(\zeta).
\label{E-intensity}
\end{equation}

\section{Comparison to Spectrum Models}
\label{S-models}

\begin{figure}[htb]
\setlength{\unitlength}{1cm}
\begin{center}
\epsfig{file=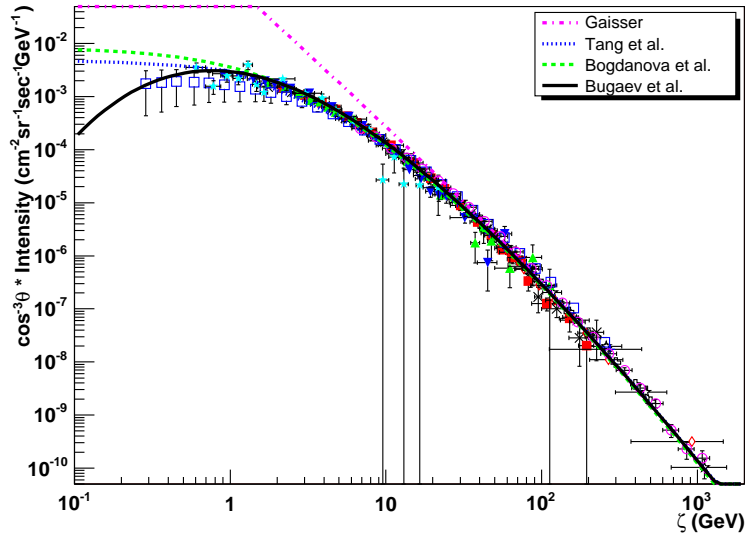, width=0.8\textwidth}
\end{center}
\caption{The combined date from Fig.~\ref{F-bestFit} are compared
to some parameterizations of vertical muon intensity (Gaisser\cite{Gaisser} 
and Bugaev\cite{Bugaev}) as well as some recent attempts at complete 
momentum and angle coverage in which the zenith angle was set to zero 
(Tang\cite{Tang} and Bogdanova\cite{Bogdanova}).} \label{F-compareFits}
\end{figure}

As a side issue, this section discusses the specific form of $I$, and
argues for the utility of the relationship expressed in Eq.~\ref{E-intensity}.
With the combined data, as shown in Fig.~\ref{F-bestFit}, it is
possible to compare several parameterizations that attempt to provide
calculations of muon intensity.  Four such models are shown in 
Fig.~\ref{F-compareFits}.

The Gaisser\cite{Gaisser} and Bugaev\cite{Bugaev} models are both 5 parameter
functions which describe only the vertical muon intensity while the
other two models (Bogdanova\cite{Bogdanova} and Tang\cite{Tang}) are
recent attempts to better match the angular dependence of the intensity 
at the surface.  The Gaisser formula is based on the physics of the muon 
production in the atmoshpere
and was validated with most of the world's data at depth.  It is not 
expected to be valid at energies below the pion threshold ($\sim$100 GeV), but
since it is a standard reference of the community, it is included here 
for comparison.

The Bugaev work is focused on nuclear cascade models for the propogation
of high-energy nucleons, pions and kaon in the atmosphere.  The formula 
provided is the result of a simple parameterization to thier calculations.

The work by Tang et al.\cite{Tang} is an attempt to improve the initial
Gaisser formula by including the effects due to the curvature of the earth 
to calculate the specific atmospheric path-length and
density corrections for all zenith angles.  In addition, an emperical
functional modification is provided for $p_\mu < 1$ GeV/$\cos\theta$ to
improve compatibility with low energy data where muon decay is expected
to be significant.

The Bogdanova model\cite{Bogdanova} is based on an optimization of the
parameters in the emperical formula first proposed by Miyake\cite{Miyake}.
It contains 3 simple terms which represent the muon production spectrum and 
the effects of pion and muon decay.

To evaluate the quality of these parameterizations, a  $\chi^2$ comparison 
of each was performed to the entire data set (shown in Table~\ref{T-chi2}).
\begin{table}[htb]
\begin{center}
\caption{The $\chi^2$ comparison between the listed models and all the 
data shown in Table~\ref{T-exp}.  In the last column, values of the 
intesity at all zenith angles are calculated by relying on the relationship
in Eq.~\ref{E-intensity} and evaluating the given models for
vertical intensity only.  For the two models that provide an alternative 
zenith angle parameterization, this is compared in the ``all $\theta$''
column, which does not rely on Eq.~\ref{E-intensity}.} \label{T-chi2}
\begin{tabular}{|l|c||c|}
\hline
Model & $\chi^2 / n.d.f.$ (all $\theta$) & 
        $\chi^2 / n.d.f.$ (use Eq.~\ref{E-intensity}) \\
\hline
Tang\cite{Tang}   & 518.69 / 191 & 354.822 / 197     \\
Bogdanova\cite{Bogdanova} & 330.235 / 203 & 387.727 / 205 \\
Bugaev\cite{Bugaev} & -- & 232.861 / 204 \\
\hline
Best Fit          & -- & 165.62 / 204 \\
\hline
\end{tabular}
\end{center}
\end{table}
Since the models of Bogdanova and Tang both provide zenith angular dependences,
they were able to be compared to the data without the scaling relationship
of Eq.~\ref{E-intensity} (shown in the column labeled ``all $\theta$'').  

However, when those same models were used for the vertical intensity only 
($\theta$ is set to zero) and the relationship expressed in 
Eq.~\ref{E-intensity} was used to predict
the angular dependance, the resulting $\chi^2$s are as good or better.
Furthermore, the parameterization provided by Bugaev et al. provides
an even better relation to the data, with a reduced $\chi^2$ approaching
one.

An attempt was made to see if the inclusion of all of the data from the various
zenith angles would allow for an improved solution of the coefficients in
the Bugaev parameterization: 
\begin{equation}
I_{\textrm v}(p_\mu) = c_1p_\mu^{-1(c_2 + c_3\log_{10}(p_\mu) + c_4\log_{10}^2(p_\mu) + c_5\log_{10}^3(p_\mu)} .
\label{E-bugaev}
\end{equation}
The original coefficients were defined separately for 4 momentum ranges: 
1 -- 927.65 GeV, 927.65 -- 1587.8 GeV, 1587.8 -- 41625 GeV, 
and $>$41625 GeV. 

Here, Eqs.~\ref{E-intensity} and \ref{E-bugaev} were combined
and the values of the parameters were allowed to float.  Fitting the 
entire sample of data, a $\chi^2$ of 165.6 for 204
degrees of freedom could be achieved (refered to as ``Best Fit'' in 
Table~\ref{T-chi2} and Fig.~\ref{F-bestFit}) -- a modest improvement -- 
by using the following coefficients:
\begin{table}[htb]
\begin{center}
\begin{tabular}{c c c c c}
$c_1$ = 0.00253 & $c_2$ = 0.2455 & $c_3$ = 1.288 & $c_4$ = -0.2555 & $c_5$ = 0.0209 \\
\end{tabular}
\end{center}
\end{table}

Recall that this improved parameterization applies to {\em all} zenith angles.
Given the available data, it can be considered valid for muon momenta 
$p_\mu >  1$ GeV and $p_\mu < 2000$ GeV/$\cos\theta$.

\section{Conclusions}

An examination of experimental data on muon intensities for various 
zenith angles at the surface has been performed.  Within that data,
a simple relationship has been found between the angular distribution
and the vertical momentum spectrum of 
$I(p_\mu,\theta) = \cos^3(\theta)I_{\textrm v}(\zeta)$ where 
$\zeta = p_\mu \cos(\theta)$.  One wonders if it is perhaps significant 
that $\zeta$ is the component of the muon momentum perpendicular to the 
surface.  Nevertheless, for the purpose of simulating muon intensities
near the surface, this relation appears remarkably accurate.

The highest accuracy was achieved when using the functional form from 
\cite{Bugaev} and adjusting the coefficients as described above (the result is
labeled ``Best Fit'' in Table~\ref{T-chi2} and Fig.~\ref{F-bestFit}).  
For simulations in which energies of a few tens of GeV are important (less
than 100 m.w.e.), it is recommended to use this modified fit.  At
greater depths, where the energies below 10 GeV can safely be neglected, 
the original parameters listed in \cite{Bugaev} would probably be preferred,
since they provide a smooth transition to higher values of $\zeta$.  Although
it was not studied in this work, it might be an interesting exercise
to validate the angular correlation of Eq.~\ref{E-intensity} at
higher energies.  

\begin{ack}
The author would like to acknowledge the assistance of Xiangyu Ding from
the Illinois Math and Science Academy for his assistance in finding 
useful data and initial efforts in data comparison and fitting.
\end{ack}

\end{document}